# MIXED-CONVECTION LAMINAR FILM CONDENSATION ON A SEMI-INFINITE VERTICAL PLATE

**Jian-Jun SHU and Graham WILKS**
School of Mechanical & Aerospace Engineering
Nanyang Technological University
50 Nanyang Avenue, Singapore 639798
E-mail: mjjshu@ntu.edu.sg

**ABSTRACT**
A comprehensive study of the problem of laminar film condensation with both a gravitational type body force and a moving vapour concurrent and parallel to the surface has been presented here. It demonstrates where both the body force and vapour velocity are significant through a comprehensive numerical solution obtained by a modified Keller box method. Important parameters governing condensation and heat transfer of pure vapour are determined. A perturbation analysis is applied in the leading edge and downstream regimes. The thin film approximations for the both regimes are obtained and compared with exact numerical solutions.

**INTRODUCTION**

The problem of laminar film condensation on a vertical plate with both a gravitational type body force and a moving vapour concurrent and parallel to the surface is not only of theoretical interest but also of great practical importance in a number of technological processes. There have been sporadic publications in the area of the laminar body-force-only film condensation problem and in the area of the laminar forced convection film condensation problem.

In the area of the laminar body-force-only film condensation, the pioneer work was reported by Nusselt (1916), who formulated this problem in terms of simple force and heat balances within the condensate film. The effects of inertia forces, energy convection and vapour drag were not taken into account. In the area of the laminar forced convection film condensation, Jacobs (1966) assumed that the inertia terms of the liquid film have negligible effect and that the temperature distribution across the film is linear. Beckett and Poots (1972) using boundary-layer similarity techniques presented a comprehensive solution to this problem without a body force.

In the present investigation, a comprehensive theoretical model for the problem of combined body force and forced convection laminar film condensation has been built up. Accordingly the heat transfer and condensation rates in this problem have accurately been predicted by employing perturbation techniques and exact numerical results which have been computed by using the continuous transformation computation (Hunt and Wilks 1981) and the modified Box scheme (Shu and Wilks 1995).

**BASIC EQUATIONS**

The flow under investigation has been modelled as a steady, two-dimensional flow of incompressible fluid. A sketch of the physical model and co-ordinate system is given in Figure 1. The flat plate is suspended in a flow of pure, saturated vapour with a gravitational field acting in the same direction as the flow which is parallel to the plate. The plate surface is maintained at a uniform temperature $T_w$ below saturation temperature $T^*$. The vapour velocity approaches the free stream vapour velocity $U$.

Let $(u,v)$ denote the condensate velocity components in the $(x,y)$-directions, where $x$ measure distance along the plate surface and $y$ distance normal to it. Let $T$ denote the temperature of the condensate. The interface separating the condensate and vapour phases is denoted by $y = \delta(x)$. For the vapour phase a set of intrinsic coordinates attached to the interface are chosen. $x^*$ measures the distance along the interface and $y^*$ the distance normal to it. $(u^*, v^*)$ denotes the velocity components of the vapour in the directions of $(x^*, y^*)$ increasing.

It is now assumed that the thickness of the condensate is small compared with a typical dimension of the surface and thus $x = x^*$. Furthermore, on the assumption that all changes in physical quantities normal to the surface (or the interface) are large compared with changes in the $x$-direction, it is permissible to invoke the boundary-layer approximation. The boundary-layer equations describing conservation of mass and momentum in both phases and thermal energy in the condensate phase are, in the usual notation, as follows.

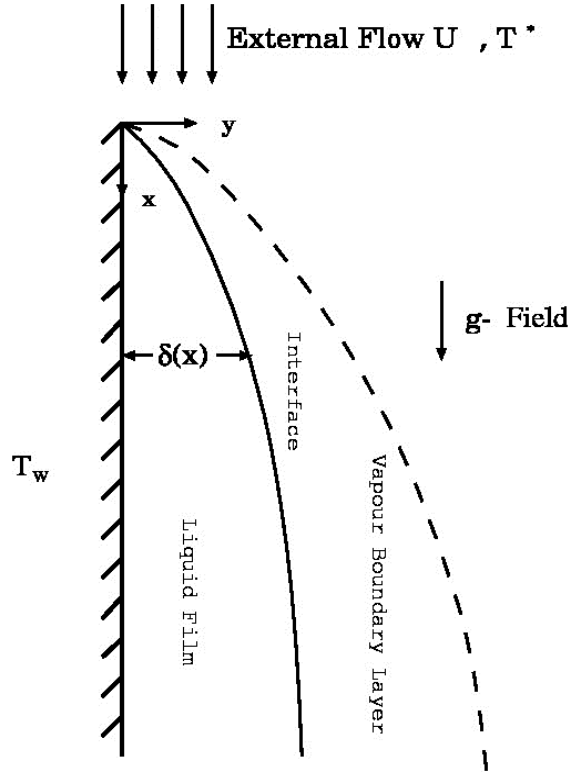

**Figure 1** Physical model and co-ordinate system.

*Condensate phase* $x \geq 0$, $0 \leq y \leq \delta(x)$,

$$\frac{\partial u}{\partial x} + \frac{\partial v}{\partial y} = 0 \tag{1}$$

$$\rho u \frac{\partial u}{\partial x} + \rho v \frac{\partial u}{\partial y} = g(\rho - \rho^*) + \mu \frac{\partial^2 u}{\partial y^2} \tag{2}$$

$$\rho C_p \left( u \frac{\partial T}{\partial x} + v \frac{\partial T}{\partial y} \right) = \kappa \frac{\partial^2 T}{\partial y^2}. \tag{3}$$

*Vapour phase* $x \geq 0$, $y^* \geq 0$, $T = T^*$,

$$\frac{\partial u^*}{\partial x} + \frac{\partial v^*}{\partial y^*} = 0 \tag{4}$$

$$\rho^* u^* \frac{\partial u^*}{\partial x} + \rho^* v^* \frac{\partial u^*}{\partial y^*} = \mu^* \frac{\partial^2 u^*}{\partial y^{*2}}. \tag{5}$$

*Boundary conditions*:

For $x \geq 0$, $y = 0$, $\quad u = 0$, $v = 0$, $T = T_w$. (6)

At the interface, $x \geq 0$, $y = \delta(x)$, $y^* = 0$, $\quad T = T^*$. (7)

Continuity in interface mass flow yields $\quad \rho\left(v - u\frac{d\delta}{dx}\right) = \rho^*\left(v^* - u^*\frac{d\delta}{dx}\right) = -\frac{d}{dx}\left(\int_0^{\delta(x)} \rho u\, dy\right)$. (8)

Continuity in tangential component of interface velocity yields $\quad u = u^*$. (9)

Continuity in interface shear stress components yields $\quad \mu \frac{\partial u}{\partial y} = \mu^* \frac{\partial u^*}{\partial y^*}$, $p = p^*$. (10)

The overall energy balance is given by $\quad -\int_0^x \kappa \frac{\partial T}{\partial y}\bigg|_{y=0} dx + \int_0^{\delta(x)} \rho u h_f\, dy + \int_0^{\delta(x)} \rho u C_P(T^* - T)\, dy = 0$. (11)

Finally, in the vapour phase $x \geq 0$, $y^* \to +\infty$, $\quad u^* \to U$. (12)

Here a star * is employed to signify a vapour quantity and $\rho$, $\mu$, $C_p$, $\kappa$ and $h_f$ denote density, dynamic viscosity, specific heat, thermal conductivity and latent heat of condensation respectively.

The significant step in formulating the problem for comprehensive solutions is the introduction of the characteristic non-dimensional co-ordinate $\xi = \dfrac{gx}{U^2}$, employed by Jacobs (1966). This co-ordinate provides the basis for a unified framework within which the features of dominant forced convection and dominant body force may be associated with small and large $\xi$ respectively.

**THIN FILM APPROXIMATION IN LEADING EDGE**

Transformations appropriate to the leading edge are introduced as

$$\psi = \lambda \sqrt{\dfrac{2\mu Ux}{\rho}} F(\xi,\phi), \qquad \phi = \dfrac{\lambda}{\eta_\delta(\xi)} \sqrt{\dfrac{\rho U}{2\mu}} \dfrac{y}{x^{\frac{1}{2}}}, \qquad (13)$$

$$\psi^* = \sqrt{\dfrac{2\mu^* Ux}{\rho^*}} f^*(\xi,\eta^*), \qquad \eta^* = \sqrt{\dfrac{\rho^* U}{2\mu^*}} \dfrac{y^*}{x^{\frac{1}{2}}}, \qquad (14)$$

$$T - T^* = (T_w - T^*)\Theta(\xi,\phi) \qquad (15)$$

where $\lambda = \left(\dfrac{\rho - \rho^*}{\rho}\right)^{\frac{1}{4}}$ and in the $(\xi,\eta)$-plane the thickness of the condensate layer is $\eta = \eta_\delta(\xi)$ when $y = \delta(x)$. The boundary-layer equations (1) - (5) and the boundary conditions (6) - (12) transform, respectively, to the following system of equations.

$$\dfrac{\partial^3 F}{\partial \phi^3} + \eta_\delta(\xi) F \dfrac{\partial^2 F}{\partial \phi^2} + 2\xi \eta_\delta^3(\xi) = 2\xi\left[\eta_\delta(\xi)\dfrac{\partial F}{\partial \phi}\dfrac{\partial^2 F}{\partial \xi \partial \phi} - \dfrac{d\eta_\delta(\xi)}{d\xi}\left(\dfrac{\partial F}{\partial \phi}\right)^2 - \eta_\delta(\xi)\dfrac{\partial^2 F}{\partial \phi^2}\dfrac{\partial F}{\partial \xi}\right] \quad 0 < \phi < 1, \qquad (16)$$

$$\dfrac{1}{P_r}\dfrac{\partial^2 \Theta}{\partial \phi^2} + \eta_\delta(\xi) F \dfrac{\partial \Theta}{\partial \phi} = 2\xi \eta_\delta(\xi)\left(\dfrac{\partial F}{\partial \phi}\dfrac{\partial \Theta}{\partial \xi} - \dfrac{\partial \Theta}{\partial \phi}\dfrac{\partial F}{\partial \xi}\right) \qquad 0 < \phi < 1, \qquad (17)$$

$$\dfrac{\partial^3 f^*}{\partial \eta^{*3}} + f^* \dfrac{\partial^2 f^*}{\partial \eta^{*2}} = 2\xi\left(\dfrac{\partial f^*}{\partial \eta^*}\dfrac{\partial^2 f^*}{\partial \xi \partial \eta^*} - \dfrac{\partial^2 f^*}{\partial \eta^{*2}}\dfrac{\partial f^*}{\partial \xi}\right) \qquad \eta^* > 0 \qquad (18)$$

with boundary conditions as follows.

At the wall $\phi = 0$, $\qquad F = 0, \quad \dfrac{\partial F}{\partial \phi} = 0, \quad \Theta = 1.$ \qquad (19)

At the interface $\phi = 1$, $\eta^* = 0$, $\quad \Theta = 0, \quad F = \dfrac{1}{\lambda \omega} f^*, \quad \dfrac{\partial F}{\partial \phi} = \dfrac{\eta_\delta(\xi)}{\lambda^2}\dfrac{\partial f^*}{\partial \eta^*}, \quad \dfrac{\partial^2 F}{\partial \phi^2} = \dfrac{\eta_\delta^2(\xi)}{\lambda^3 \omega}\dfrac{\partial^2 f^*}{\partial \eta^{*2}},$ \qquad (20)

$$H_0 \dfrac{\partial \Theta}{\partial \phi} + \eta_\delta(\xi) F + 2\xi \eta_\delta(\xi)\dfrac{\partial F}{\partial \xi} = 0. \qquad (21)$$

In the vapour as $\eta^* \to +\infty$, $\qquad \dfrac{\partial f^*}{\partial \eta^*} \to 1$ \qquad (22)

where $\omega = \left(\dfrac{\rho\mu}{\rho^*\mu^*}\right)^{\frac{1}{2}}$ and $H_0 = \dfrac{C_p(T^* - T_w)}{P_r h_f}$. The following quantities are of practical interest: Prandtl number $P_r = \dfrac{C_p \mu}{\kappa}$, Nusselt number $N_u = \dfrac{x}{T^* - T_w}\dfrac{\partial T}{\partial y}\bigg|_{y=0}$, Reynolds number $R_e = \dfrac{\rho U x}{\mu}$ and skin friction coefficient $C_f = \dfrac{2\mu}{\rho U^2}\dfrac{\partial u}{\partial y}\bigg|_{y=0}$.

The thin film is specified by either of the inequalities $\dfrac{T^* - T_w}{T^*} \ll 1$ or $H_0 \ll 1$. To fit this requirement it is necessary to expand the velocity and thermal fields in terms of the parameter $\varepsilon = H_0^{\frac{1}{3}}$. It is the form of the equations which indicates the appropriateness of a regular perturbation scheme in powers of $\xi$ as

$$F(\xi,\phi) = \sum_{n=0}^{+\infty}\sum_{m=2}^{+\infty} \xi^n \varepsilon^m F_{nm}(\phi), \quad \Theta(\xi,\phi) = \sum_{n=0}^{+\infty}\sum_{m=0}^{+\infty} \xi^n \varepsilon^m \Theta_{nm}(\phi), \quad f^*(\xi,\eta^*) = \sum_{n=0}^{+\infty}\sum_{m=0}^{+\infty} \xi^n \varepsilon^m f_{nm}^*(\eta^*)$$

and

$$\eta_\delta(\xi) = \sum_{n=0}^{+\infty}\sum_{m=1}^{+\infty} \dfrac{\xi^n \varepsilon^m}{n!}\dfrac{d^n \eta_m}{d\xi^n}\bigg|_{\xi=0}$$

where $\eta_\delta(\xi)$ is expressed in form of the Taylor expansion. After some manipulation, the approximate solutions of the equations are

$$F(\xi,\phi) = \frac{1}{\lambda}\left(\frac{A_0 H_0^2}{2\omega}\right)^{\frac{1}{3}}\phi^2 - \frac{2\lambda^3\omega\xi H_0}{21A_0}\phi^2(7\phi-9) + O\left(\xi^2 + H_0^{\frac{4}{3}}\right) \tag{23}$$

$$\Theta(\xi,\phi) = 1 - \phi + \frac{P_r H_0}{12}\phi(\phi^3 - 1) - \frac{P_r\lambda\xi H_0^{\frac{4}{3}}}{60}\left(\frac{2\omega}{A_0}\right)^{\frac{1}{3}}\phi(3\phi^4 - 5\phi^3 + 2) + O\left(\xi^2 + H_0^{\frac{5}{3}}\right) \tag{24}$$

$$f^*(\xi,\eta^*) = G(\eta^*) + \left(\frac{2H_0}{A_0\omega^2}\right)^{\frac{1}{3}}\frac{dG}{d\eta^*} + O\left(\xi^2 + H_0^{\frac{2}{3}}\right) \tag{25}$$

$$\eta_\delta(\xi) = \lambda\left(\frac{2\omega H_0}{A_0}\right)^{\frac{1}{3}} - \frac{2\lambda^5\xi H_0^{\frac{2}{3}}}{7}\left(\frac{2\omega}{A_0}\right)^{\frac{5}{3}} + O(\xi^2 + H_0) \tag{26}$$

where $A_0 = \left.\frac{d^2 G}{d\eta^{*2}}\right|_{\eta^*=0} = 0.4696$ and $G(\eta^*)$ satisfies

$$\frac{d^3 G}{d\eta^{*3}} + G\frac{d^2 G}{d\eta^{*2}} = 0 \quad \text{with} \quad G|_{\eta^*=0} = \left.\frac{dG}{d\eta^*}\right|_{\eta^*=0} = 0, \quad \left.\frac{dG}{d\eta^*}\right|_{\eta^*\to+\infty} \to 1. \tag{27}$$

The flow characteristics are given by

$$\frac{N_u}{\sqrt{R_e}} = \frac{1}{\sqrt{2}}\left(\frac{A_0}{2\omega H_0}\right)^{\frac{1}{3}}\left[1 + \frac{2\lambda^4\xi H_0^{\frac{1}{3}}}{7}\left(\frac{2\omega}{A_0}\right)^{\frac{4}{3}} + O\left(\xi^2 + H_0^{\frac{2}{3}}\right)\right], \tag{28}$$

$$C_f\sqrt{R_e} = \frac{\sqrt{2}A_0}{\omega}\left[1 + \lambda^4\xi H_0^{\frac{1}{3}}\left(\frac{2\omega}{A_0}\right)^{\frac{4}{3}} + O\left(\xi^2 + H_0^{\frac{2}{3}}\right)\right], \tag{29}$$

$$\frac{\delta(x)\sqrt{R_e}}{x} = \sqrt{2}\left(\frac{2\omega H_0}{A_0}\right)^{\frac{1}{3}}\left[1 - \frac{2\lambda^4\xi H_0^{\frac{1}{3}}}{7}\left(\frac{2\omega}{A_0}\right)^{\frac{4}{3}} + O\left(\xi^2 + H_0^{\frac{2}{3}}\right)\right]. \tag{30}$$

These expressions are only applicable when the inflow velocity at the interface is small. This implies the inequality $\omega H_0 \ll \sqrt{\frac{2}{A_0}} \approx 2.0637$, or in the case of steam-water condensation that $T^* - T_w \ll 10^\circ C$.

**THIN FILM APPROXIMATION IN DOWNSTREAM REGIME**

Transformations appropriate to the asymptotic expansion are introduced as

$$\psi = \left[\frac{64g(\rho-\rho^*)\mu^2 x^3}{\rho^3}\right]^{\frac{1}{4}}\overline{F}(\xi,\overline{\phi}), \quad \overline{\phi} = \frac{1}{\overline{\eta}_\delta(\xi)}\left[\frac{g\rho(\rho-\rho^*)}{4\mu^2}\right]^{\frac{1}{4}}\frac{y}{x^{\frac{1}{4}}}, \tag{31}$$

$$\psi^* = \frac{\mu\omega}{\rho}\left(\frac{64g\rho^{*2}x^3}{\mu^{*2}}\right)^{\frac{1}{4}}\overline{F}^*(\xi,\overline{\phi}^*), \quad \overline{\phi}^* = \omega\left(\frac{g\rho^{*2}}{4\mu^{*2}}\right)^{\frac{1}{4}}\frac{y^*}{x^{\frac{1}{4}}}, \tag{32}$$

$$T - T^* = (T_w - T^*)\overline{\Theta}(\xi,\overline{\phi}). \tag{33}$$

In the $(\xi,\overline{\eta})$-plane the thickness of the condensate layer is $\overline{\eta} = \overline{\eta}_\delta(\xi)$ when $y = \delta(x)$. The boundary-layer equations (1) - (5) and the boundary conditions (6) - (12) transform, respectively, to the following system of equations.

$$\frac{\partial^3\overline{F}}{\partial\overline{\phi}^3} + \overline{\eta}_\delta(\xi)\left[3\overline{F}\frac{\partial^2\overline{F}}{\partial\overline{\phi}^2} - 2\left(\frac{\partial^2\overline{F}}{\partial\overline{\phi}^2}\right)^2\right] + \overline{\eta}_\delta^3(\xi) = 4\xi\left[\overline{\eta}_\delta(\xi)\frac{\partial\overline{F}}{\partial\overline{\phi}}\frac{\partial^2\overline{F}}{\partial\xi\partial\overline{\phi}} - \frac{d\overline{\eta}_\delta(\xi)}{d\xi}\left(\frac{\partial\overline{F}}{\partial\overline{\phi}}\right)^2 - \overline{\eta}_\delta(\xi)\frac{\partial^2\overline{F}}{\partial\overline{\phi}^2}\frac{\partial\overline{F}}{\partial\xi}\right] \quad 0 < \overline{\phi} < 1, \tag{34}$$

$$\frac{1}{P_r}\frac{\partial^2\overline{\Theta}}{\partial\overline{\phi}^2} + 3\overline{\eta}_\delta(\xi)\overline{F}\frac{\partial\overline{\Theta}}{\partial\overline{\phi}} = 4\xi\overline{\eta}_\delta(\xi)\left(\frac{\partial\overline{F}}{\partial\overline{\phi}}\frac{\partial\overline{\Theta}}{\partial\xi} - \frac{\partial\overline{\Theta}}{\partial\overline{\phi}}\frac{\partial\overline{F}}{\partial\xi}\right) \quad 0 < \overline{\phi} < 1, \tag{35}$$

$$\frac{\partial^3\overline{F}^*}{\partial\overline{\phi}^{*3}} + 3\overline{F}^*\frac{\partial^2\overline{F}^*}{\partial\overline{\phi}^{*2}} - 2\left(\frac{\partial\overline{F}^*}{\partial\overline{\phi}^*}\right)^2 = 4\xi\left(\frac{\partial\overline{F}^*}{\partial\overline{\phi}^*}\frac{\partial^2\overline{F}^*}{\partial\xi\partial\overline{\phi}^*} - \frac{\partial^2\overline{F}^*}{\partial\overline{\phi}^{*2}}\frac{\partial\overline{F}^*}{\partial\xi}\right) \quad \overline{\phi}^* > 0 \tag{36}$$

with boundary conditions

At the wall $\bar{\phi} = 0$,
$$\bar{F} = 0, \quad \frac{\partial \bar{F}}{\partial \bar{\phi}} = 0, \quad \bar{\Theta} = 1. \tag{37}$$

At the interface $\bar{\phi} = 1$, $\bar{\phi}^* = 0$, $\quad \bar{\Theta} = 0, \quad \bar{F} = \frac{1}{\lambda}\bar{F}^*, \quad \frac{\partial \bar{F}}{\partial \bar{\phi}} = \frac{\omega^2 \bar{\eta}_\delta(\xi)}{\lambda^2} \frac{\partial \bar{F}^*}{\partial \bar{\phi}^*}, \quad \frac{\partial^2 \bar{F}}{\partial \bar{\phi}^2} = \frac{\omega^2 \bar{\eta}_\delta^2(\xi)}{\lambda^3} \frac{\partial^2 \bar{F}^*}{\partial \bar{\phi}^{*2}}, \tag{38}$

$$H_0 \frac{\partial \bar{\Theta}}{\partial \bar{\phi}} + 3\bar{\eta}_\delta(\xi)\bar{F} + 4\xi\bar{\eta}_\delta(\xi)\frac{\partial \bar{F}}{\partial \xi} = 0. \tag{39}$$

In the vapour as $\bar{\phi}^* \to +\infty$,
$$\frac{\partial \bar{F}^*}{\partial \bar{\phi}^*} \to \frac{1}{2}\omega^{-2}\xi^{-\frac{1}{2}}. \tag{40}$$

Although the thin film is specified by either of the inequalities $\frac{T^* - T_w}{T^*} \ll 1$ or $H_0 \ll 1$, the solution of the equations (34)-(40) does not exist when $\omega = O(1)$. In fact the inequality $\omega \gg 1$ is applicable for most vapours, for example, in the case of steam-water condensation $\omega = 191$ for $0^\circ C \le T^* - T_w \le 100^\circ C$. Since $\omega$ is large it is necessary to expand the velocity and thermal fields in terms of the parameter $\omega^{-2}$. It is the form of the last boundary condition which suggests in the first instance the appropriate perturbation scheme as

$$\bar{F}(\xi, \bar{\phi}) = H_0^{\frac{3}{4}} \bar{F}_0(\bar{\phi}) + O(\xi^{-1}\ln\xi + \omega^{-2} + H_0),$$

$$\bar{\Theta}(\xi, \bar{\phi}) = \bar{\Theta}_0(\bar{\phi}) + O\left(\xi^{-1}\ln\xi + \omega^{-2} + H_0^{\frac{1}{4}}\right),$$

$$\frac{\partial \bar{F}^*}{\partial \bar{\phi}^*} = \omega^{-2}\left[\bar{F}_0^*(\bar{\phi}^*) + \xi^{-\frac{1}{2}}\bar{F}_1^*(\bar{\phi}^*)\right] + O(\xi^{-1}\ln\xi + \omega^{-4})$$

and

$$\bar{\eta}_\delta(\xi) = H_0^{\frac{1}{4}}\bar{\eta}_0 + O\left(\xi^{-1}\ln\xi + \omega^{-2} + H_0^{\frac{1}{2}}\right).$$

After some manipulation, the approximate solutions of the equations are

$$\bar{F}(\xi, \bar{\phi}) = -\frac{1}{6}H_0^{\frac{3}{4}}\bar{\phi}^2(\bar{\phi} - 3) + O(\xi^{-1}\ln\xi + \omega^{-2} + H_0), \tag{41}$$

$$\bar{\Theta}(\xi, \bar{\phi}) = 1 - \bar{\phi} + O\left(\xi^{-1}\ln\xi + \omega^{-2} + H_0^{\frac{1}{4}}\right), \tag{42}$$

$$\frac{\partial \bar{F}^*}{\partial \bar{\phi}^*} = \omega^{-2}\left\{\frac{1}{2}\lambda^2 H_0^{\frac{1}{2}} + \frac{1}{16}\xi^{-\frac{1}{2}}\left[8 - \left(8 - 6H_0 + \lambda H_0^{\frac{7}{4}}\bar{\phi}^*\right)\right]\right\}\exp\left(-\lambda H_0^{\frac{3}{4}}\bar{\phi}^*\right) + O(\xi^{-1}\ln\xi + \omega^{-4}), \tag{43}$$

$$\bar{\eta}_\delta(\xi) = H_0^{\frac{1}{4}} + O\left(\xi^{-1}\ln\xi + \omega^{-2} + H_0^{\frac{1}{2}}\right). \tag{44}$$

The flow characteristics are given by

$$\frac{N_u}{\sqrt{R_e}} = \frac{\lambda}{\sqrt{2}}\left(\frac{\xi}{H_0}\right)^{\frac{1}{4}}\left[1 + O\left(\xi^{-1}\ln\xi + \omega^{-2} + H_0^{\frac{1}{4}}\right)\right], \tag{45}$$

$$C_f\sqrt{R_e} = 2\sqrt{2}\lambda^3 H_0^{\frac{1}{4}}\xi^{\frac{3}{4}}\left[1 + O\left(\xi^{-1}\ln\xi + \omega^{-2} + H_0^{\frac{1}{4}}\right)\right], \tag{46}$$

$$\frac{\delta(x)\sqrt{R_e}}{x} = \frac{\sqrt{2}}{\lambda}\left(\frac{H_0}{\xi}\right)^{\frac{1}{4}}\left[1 + O\left(\xi^{-1}\ln\xi + \omega^{-2} + H_0^{\frac{1}{4}}\right)\right]. \tag{47}$$

## NUMERICAL SOLUTIONS

The perturbation solutions at small $\xi$ and large $\xi$, the allied similarity solutions and thin film approximations have been obtained. They provide some useful local information for dominant forced convection and dominant body force, but it is clear that there exists an extensive transitional region of $\xi$ for which the respective solutions at small $\xi$ and large $\xi$ are inadequate for precise estimates of the heat transfer characteristics. Such detailed information may only be obtained by a comprehensive numerical solution of the governing system of equations (1)-(5) under boundary conditions (6)-(12).

Hunt and Wilks (1981) have demonstrated the advantages of introducing a continuous transformation in the characterizing co-ordinate. It is this method which is adapted to the present problem to yield a single set of equations which adequately accommodate the essential features of each of the two extreme regimes and accordingly provide a unified basis upon which complete numerical solution may be obtained.

The following transformations are introduced

$$\psi = \lambda\sqrt{\frac{2\mu U x}{\rho}}r(\xi)\tilde{f}(\xi,\tilde{\eta}), \qquad \tilde{\eta} = \lambda\sqrt{\frac{\rho U}{2\mu}}\frac{y}{x^{\frac{1}{2}}}t(\xi), \qquad (48)$$

$$\psi^* = \sqrt{\frac{2\mu^* U x}{\rho^*}}r^*(\xi)\tilde{f}^*(\xi,\tilde{\eta}^*), \qquad \tilde{\eta}^* = \sqrt{\frac{\rho^* U}{2\mu^*}}\frac{y^*}{x^{\frac{1}{2}}}t^*(\xi), \qquad (49)$$

$$T - T^* = (T_w - T^*)s(\xi)\tilde{\theta}(\xi,\tilde{\eta}) \qquad (50)$$

where $r(\xi)$, $t(\xi)$, $r^*(\xi)$, $t^*(\xi)$ and $s(\xi)$ are to be chosen to effect a smooth transition between the two extreme regimes. Under these transformations the governing equations (1)-(5) and boundary conditions (6)-(12) become

$$\frac{\partial^3 \tilde{f}}{\partial \tilde{\eta}^3} + \frac{1}{rt}\frac{d(\xi r^2)}{d\xi}\tilde{f}\frac{\partial^2 \tilde{f}}{\partial \tilde{\eta}^2} - \frac{2\xi}{t^2}\frac{d(rt)}{d\xi}\left(\frac{\partial \tilde{f}}{\partial \tilde{\eta}}\right)^2 + \frac{2\xi}{t^3 r} + \frac{2\xi r}{t}\left(\frac{\partial \tilde{f}}{\partial \xi}\frac{\partial^2 \tilde{f}}{\partial \tilde{\eta}^2} - \frac{\partial \tilde{f}}{\partial \tilde{\eta}}\frac{\partial^2 \tilde{f}}{\partial \xi \partial \tilde{\eta}}\right) = 0, \qquad (51)$$

$$\frac{\partial^2 \tilde{\theta}}{\partial \tilde{\eta}^2} + P_r\left[\frac{1}{rt}\frac{d(\xi r^2)}{d\xi}\tilde{f}\frac{\partial \tilde{\theta}}{\partial \tilde{\eta}} - \frac{2\xi r}{st}\frac{ds}{d\xi}\frac{\partial \tilde{f}}{\partial \tilde{\eta}}\tilde{\theta} + \frac{2\xi r}{t}\left(\frac{\partial \tilde{f}}{\partial \xi}\frac{\partial \tilde{\theta}}{\partial \tilde{\eta}} - \frac{\partial \tilde{f}}{\partial \tilde{\eta}}\frac{\partial \tilde{\theta}}{\partial \xi}\right)\right] = 0, \qquad (52)$$

$$\frac{\partial^3 \tilde{f}^*}{\partial \tilde{\eta}^{*3}} + \frac{1}{r^*t^*}\frac{d(\xi r^{*2})}{d\xi}\tilde{f}^*\frac{\partial^2 \tilde{f}^*}{\partial \tilde{\eta}^{*2}} - \frac{2\xi}{t^{*2}}\frac{d(r^*t^*)}{d\xi}\left(\frac{\partial \tilde{f}^*}{\partial \tilde{\eta}^*}\right)^2 + \frac{2\xi r^*}{t^*}\left(\frac{\partial \tilde{f}^*}{\partial \xi}\frac{\partial^2 \tilde{f}^*}{\partial \tilde{\eta}^{*2}} - \frac{\partial \tilde{f}^*}{\partial \tilde{\eta}^*}\frac{\partial^2 \tilde{f}^*}{\partial \xi \partial \tilde{\eta}^*}\right) = 0 \qquad (53)$$

with boundary conditions

At the wall $\tilde{\eta} = 0$, $\qquad \tilde{f} = 0$, $\frac{\partial \tilde{f}}{\partial \tilde{\eta}} = 0$, $s\tilde{\theta} = 1$. $\qquad (54)$

At the interface $\tilde{\eta} = \tilde{\eta}_\delta(\xi)$, $\tilde{\eta}^* = 0$, $\qquad \tilde{\theta} = 0$, $\tilde{f} = \frac{r^*}{\lambda \omega r}\tilde{f}^*$, $\frac{\partial \tilde{f}}{\partial \tilde{\eta}} = \frac{r^*t^*}{\lambda^2 rt}\frac{\partial \tilde{f}^*}{\partial \tilde{\eta}^*}$, $\frac{\partial^2 \tilde{f}}{\partial \tilde{\eta}^2} = \frac{r^*t^{*2}}{\lambda^3 \omega rt^2}\frac{\partial^2 \tilde{f}^*}{\partial \tilde{\eta}^{*2}}$, $\qquad (55)$

$$H_0\left\{\frac{\partial \tilde{\theta}}{\partial \tilde{\eta}}\bigg|_{\tilde{\eta}=0} + \frac{P_r}{t}\left[\frac{1}{r}\frac{d(r^2\xi)}{d\xi} + \frac{2r\xi}{s}\frac{ds}{d\xi}\right]\int_0^{\tilde{\eta}_\delta(\xi)}\frac{\partial \tilde{f}}{\partial \tilde{\eta}}\tilde{\theta}d\tilde{\eta} + \frac{2P_r r\xi}{t}\int_0^{\tilde{\eta}_\delta(\xi)}\left(\frac{\partial^2 \tilde{f}}{\partial \xi \partial \tilde{\eta}}\tilde{\theta} + \frac{\partial \tilde{f}}{\partial \tilde{\eta}}\frac{\partial \tilde{\theta}}{\partial \xi}\right)d\tilde{\eta}\right\} + \frac{\tilde{f}}{rst}\frac{d(r^2\xi)}{d\xi} + \frac{2r\xi}{st}\left(1 + \frac{d\tilde{\eta}_\delta}{d\xi}\right)\frac{\partial \tilde{f}}{\partial \xi} = 0 \qquad (56)$$

In the vapour as $\tilde{\eta}^* \to +\infty$, $\qquad r^*t^*\dfrac{\partial \tilde{f}^*}{\partial \tilde{\eta}^*} \to 1$. $\qquad (57)$

Accordingly the formulas for the flow characteristics are

$$\frac{N_u}{\sqrt{R_e}} = -\frac{\lambda s(\xi)t(\xi)}{\sqrt{2}}\frac{\partial \tilde{\theta}}{\partial \tilde{\eta}}\bigg|_{\tilde{\eta}=0}, \qquad (58)$$

$$C_f\sqrt{R_e} = \sqrt{2}\lambda^3 r(\xi)t^2(\xi)\frac{\partial^2 \tilde{f}}{\partial \tilde{\eta}^2}\bigg|_{\tilde{\eta}=0}. \qquad (59)$$

Finally the condensate thickness yields

$$\frac{\delta(x)\sqrt{R_e}}{x} = \frac{\sqrt{2}}{\lambda t(\xi)}\tilde{\eta}_\delta(\xi). \qquad (60)$$

In the $(\xi,\tilde{\eta})$-plane the thickness of the condensate layer is $\tilde{\eta} = \tilde{\eta}_\delta(\xi)$ when $y = \delta(x)$. It is clear that the transformations are allied to the two extreme leading-edge and downstream regimes. The simplest forms are to be chosen as

$$r(\xi) = r^*(\xi) = (1 + 16\xi)^{\frac{1}{4}}, \quad s(\xi) = 1, \quad t(\xi) = t^*(\xi) = (1 + \xi)^{\frac{1}{4}}. \qquad (61)$$

It has been shown that the heat transfer characteristics depend on the five physical parameters, $\xi$, $P_r$, $H_0$, $\lambda$ and $\omega$. To facilitate comparison the dimensionless flow characteristics evaluated from (58)-(60) are plotted against $\left(\dfrac{\xi}{1+\xi}\right)^{\frac{1}{4}}$ in Figure 2. The differences at $\xi = 0$ are associated with the approximations of thin film theory as opposed to the exact numerical solution of the full equations. In the present case $\tilde{\eta}_\delta(0) = 0.66129$ is a moderate film thickness rather than thin.

It emerges, from (26) and (44), that the condensate thickness varies in proportion to $H_0^{\frac{1}{3}}$ for the leading edge and $H_0^{\frac{1}{4}}$ for the downstream regimes. The local heat transfer at the surface of plate is independent of $\omega$.

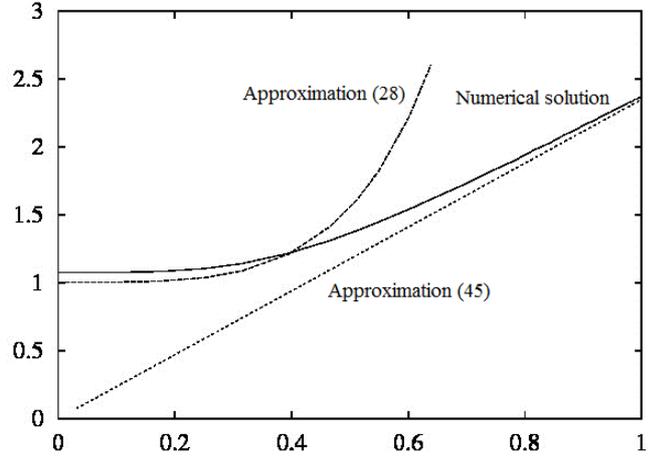

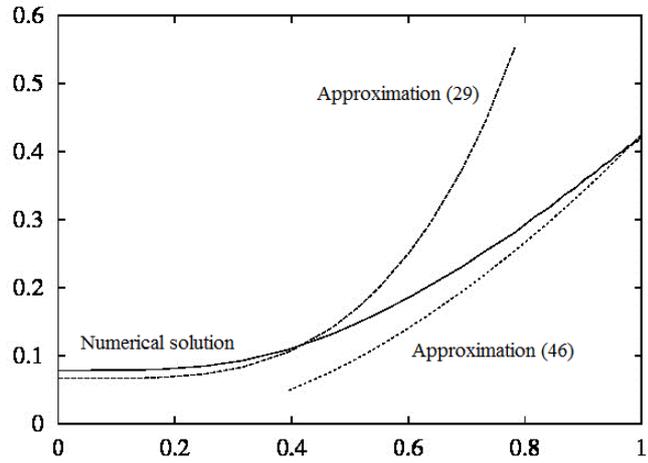

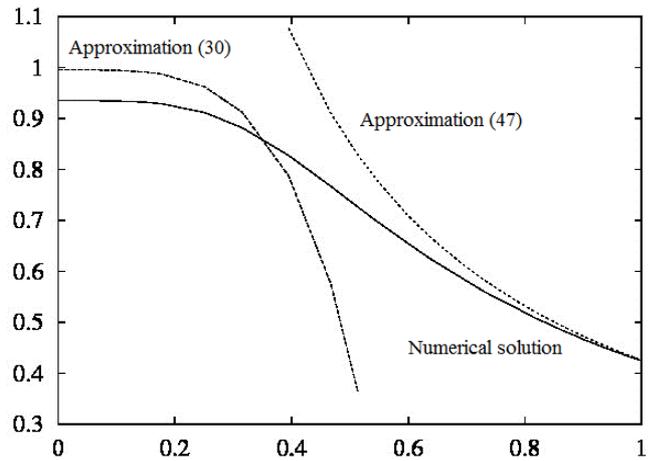

**Figure 2** Variation of (a) heat transfer $\dfrac{N_u}{(1+\xi)^{\frac{1}{4}}\sqrt{R_e}}$, (b) skin friction $\dfrac{C_f \sqrt{R_e}}{(1+16\xi)^{\frac{1}{4}}(1+\xi)^{\frac{1}{2}}}$ and (c) condensate thickness $\dfrac{(1+\xi)^{\frac{1}{4}}\delta(x)\sqrt{R_e}}{x}$ against $\left(\dfrac{\xi}{1+\xi}\right)^{\frac{1}{4}}$ for $P_r = 10$, $H_0 = 0.008191$, $\lambda = 1$ and $\omega = 10$.

Though the effect of $P_r$ was checked here only for the particular values of the parameters, it is reasonable to believe that the solution of the present problem does not depend on $P_r$ explicitly except for extreme values of the Prandtl number. The effect of the Prandtl number appears through the parameter $H_0$ only and $\lambda$ has been set to *1* as $\rho \gg \rho^*$ for most vapours. Therefore the major governing parameters are $\xi$, the ratio of the temperature potential to the heat of condensation $H_0$ and the ratio of the product of the density and viscosity of the liquid to that of vapour $\omega$.

The numerical solutions have been obtained and the results for the heat transfer are illustrated in Figure 3. With $T^* - T_w$ at values *10° F*, *40° F* and *70° F* the corresponding values of $H_0$ are *0.00617*, *0.02468* and *0.04319*. The value $\omega = 100$ was incorporated which coincides with the value chosen as the basis for presenting the experimental results. Agreement is quite convincing especially for thin films *i.e.* lower values of $H_0$.

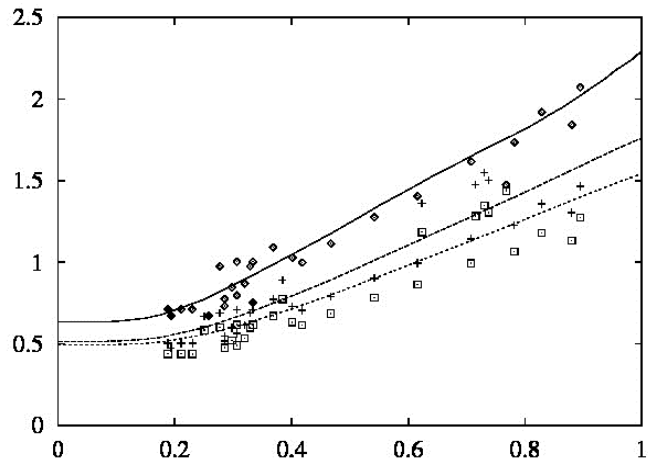

**Figure 3** Comparison of numerical and experimental results of heat transfer $\dfrac{N_u}{(1+\xi)^{\frac{1}{4}} \sqrt{R_e}}$ against $\left(\dfrac{\xi}{1+\xi}\right)^{\frac{1}{4}}$ for various physical parameter $H_0$ (◊ = *0.00617*, + = *0.02468* and □ = *0.04319*) at $P_r = 1$, $\lambda = 1$ and $\omega = 100$ for Freon 113.

## CONCLUSIONS

The combined forced convection-body force laminar film condensation problem has been examined by perturbation and numerical methods. The limiting similarity states have provided the framework for the numerical algorithm. As a consequence accurate solutions over the full range of $\xi$ have been obtained. In view of the limited range of validity of the perturbation solutions the numerical solution is essential if comprehensive heat transfer and film characteristics are to be evaluated. Comparison with experiment has been seen to be very good, particularly in the thin film setting (Shu 2004, Shu and Pop 1997, 1998a,b, 1999, Shu and Wilks 1996, 2008, 2009).